\documentclass[aps, twocolumn]{revtex4}
\usepackage{float}
\usepackage{dcolumn}
\usepackage{amsmath}
\input{epsf}
\input{epsfx}

\ifx\pdftexversion\undefined
  \usepackage{graphicx}
\else
  \usepackage[pdftex]{graphicx}
\fi

\begin{document}
\title{Efficient Photonic Crystal Cavity-Waveguide Couplers}


\author{Andrei Faraon, Dirk Englund, Ilya Fushman, Jelena Vu\v{c}kovi\'{c}}
\affiliation{E. L. Ginzton Laboratory, Stanford University, Stanford, CA, 94305}
\author{Edo Waks}
\affiliation{Department of Electrical and Computer Engineering, University of Maryland, College Park, MD, 20742}

\date{October 12, 2006}
%

\begin{abstract}

Coupling of photonic crystal (PC) linear three-hole defect cavities (L3) to PC waveguides is theoretically and experimentally investigated. The systems are designed to increase the overlap between the evanescent cavity field and the waveguide mode, and to operate in the linear dispersion region of the waveguide. Our simulations indicate increased coupling when the cavity is tilted by {$ 60 ^o $} with respect to the waveguide axis, which we have also confirmed by experiments. We obtained up to 90\% coupling efficiency into the waveguide. 

\end{abstract}

\maketitle

Structures that consist of InGaAs/GaAs quantum dots (QDs) coupled to two-dimensional PC cavities are promising candidates for highly efficient single photon sources. They represent essential devices for quantum cryptography and quantum computation  \cite{Vuckovic03APL, EdoDIT, 05PRLEnglund, VuckovicSingPhotDem}. Efficient implementation of quantum computation devices requires integration of photonic circuits directly on the chip. These circuits consist of single photon sources (SPSs) that inject single photons into the waveguides, which subsequently redirect them to other quantum nodes, i.e. other PC cavities containing QDs. Once the necessary quantum operations have been performed, photons need to be outcoupled from the waveguide either out-of-plane for vertical collection (e.g., by coupling the photons back into an ``output cavity" that scatters them out of plane), or collected in the PC plane (e.g., by outcoupling to a fiber). The performance of this kind of circuit is limited by the coupling efficiency between the cavities and the waveguides. Our work investigates this coupling with the goal of improving the efficiency of single photon transmission from one cavity to another. The results are also relevant for channel drop filter applications in optical telecommunications.

In this paper we investigate the coupling of linear three-hole cavities (L3) \cite{NodaL3} into PC waveguides. We choose the L3 cavities for their high quality factor(Q) to mode volume (V) ratio and good matching between cavity and waveguide field patterns, which improves in-plane coupling efficiency \cite{EFV05OptExpr, EdoCoupledModeTh}. The cavity mode we work with has  magnetic field with even/odd symmetry with respect to the {\bf x/y} axes. This mode, whose magnetic field configuration is depicted in Fig. 1(a), needs to be coupled to one of the guided modes in the PC waveguide. The field is computed using three-dimensional finite difference time domain simulations (3D FDTD). Of the possible waveguide bands inside the PC band gap \cite{EFV05OptExpr} the best choice for coupling the L3 cavity is the one with similar symmetry and frequency as the L3 cavity mode (Fig. 1(b)). To get efficient coupling, the cavity and waveguide modes need to be spatially overlapped and frequency matched. A closer look at the L3 cavity field profile (Fig. 1(a)) reveals that the evanescent field is strongest along a direction tilted with respect to the cavity axis and is weak along the cavity axis. A good approach for obtaining a larger overlap between the cavity and waveguide mode is to tilt the cavity with respect to the waveguide axis by an angle of {$ 60 ^{o} $} (Fig. 1(c)). The choice of this angle is determined by the symmetry constraints of the triangular lattice. Directional couplers with cavity axes non-parallel to waveguide axes have recently been studied by Kim et al \cite{NotomiHex} for coupling the hexapole modes of single hole defect cavities and by Shinya et al for coupling L3 and L4 cavities \cite{NotomiFlipFlop}. In contrast with previous work, we present here optimized designs of couplers as well as detailed theoretical and experimental data, confirming the advantage of the tilted configuration for coupling L3 cavities to PC waveguides.

\begin{figure}[htbp]
    \includegraphics[width=3in]{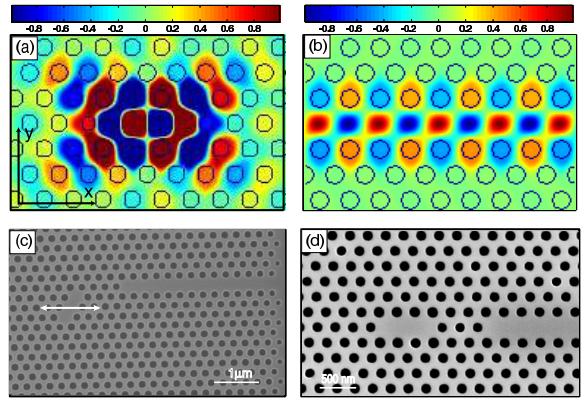}
    \caption{ (a) Magnetic field ($ B _z $ component) for the mode with the highest quality factor in a L3 cavity. (b) Magnetic field pattern of the even mode in a PC waveguide. (c) Fabricated tilted cavity coupled to a waveguide (four holes separation). In this experiment we shift the cavity with respect to the waveguide along the direction indicated by the arrow. (d) Fabricated straight cavity coupled to a waveguide (three holes separation).}
    \label{figureone}
\end{figure}

To test the validity of our approach, we compare the coupling parameters for the tilted cavity configuration (Fig. 1(c)) to the standard approach where the cavity and the waveguide share the same axis (straight cavity configuration) (Fig. 1(d)). First, 3D FDTD simulations of coupled cavity waveguide systems were performed with both tilted and straight couplers. The frequency of the waveguide band was lowered with respect to the cavity frequency by reducing the size of the PC holes that bound the waveguide. In this way, coupling occurs in the dispersion-free linear region of the waveguide band. We directly simulated tilted and straight coupler configurations with spacing of two-to-five lattice holes separation between the cavity and the waveguide. An image of the simulated magnetic field profile for a tilted cavity coupled to a waveguide with three-hole separation is depicted in Fig. 2(inset) . In the tilted configuration, the separation between the cavity and the waveguide is changed along a direction indicated by the arrow in Fig. 1(c).

\begin{figure}[htbp]
    \includegraphics[width=3in]{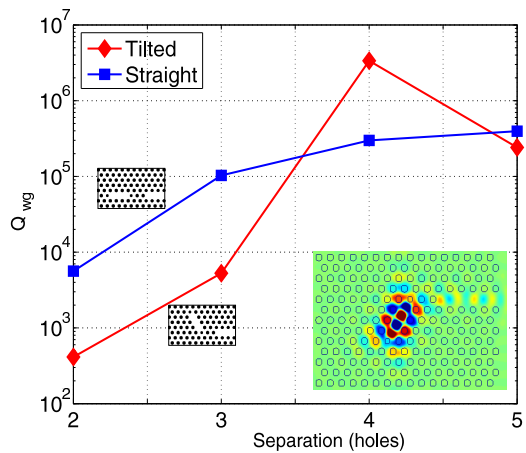}
    \caption{ Simulation results for the cavity waveguide coupling expressed in terms of the quality factor. The coupling strength is proportional to ({$ 1/Q _{wg}$}). Simulated magnetic field of a cavity-waveguide coupler in tilted configuration with three hole separation (inset). }
\end{figure}

The energy transfer into the waveguide degrades the Q of the coupled cavity. The total Q of a coupled cavity relates to the uncoupled cavity quality factor ({$ Q _{c}$}) according to:

\begin {equation} {Q _{tot}} ^{-1} = {Q _{c}}^{-1} + {Q _{wg}} ^{-1}, \end {equation}

\noindent where {$ Q _{wg} ^{-1} $} is the loss rate into the waveguide.

Different applications require different coupling. For high-efficiency single photon transfer, the in-plane coupling into the waveguide modes needs to be dominant so {$ Q _{wg}$} should be lower than {$ Q _{c} $}. On the other hand, the advanced single photon sources \cite{05PRLEnglund} require cavities with a quality factor on the order of thousands, which implies {$ Q _{wg}$} should also be in the same range. For other applications, single photons need to be scattered out of plane from a PC waveguide through an output cavity. To achieve high transfer efficiency from waveguides to the output cavities, the cavity-waveguide system needs to be in the critical coupling regime defined by {$ Q _{wg} = Q _{c} $}.  In that case, we do not need the output cavity to have a high quality factor.

The coupling strength between the cavity and the waveguide is given by {$ 1/Q _{wg} $} which is proportional to the decay rate of the cavity field into the waveguide. The quality factor {$ Q _{wg} $} was computed from the 3D FDTD simulations, with results presented in Fig. 2. For the same cavity-waveguide separation, {$ Q _{wg} $} is generally smaller for the tilted than for the straight configuration. This is an indication of better cavity-waveguide coupling obtained by tilting the cavity. One peculiar aspect of the simulations is that for the tilted coupling configuration, the Q is actually larger for four-holes than for five-holes separation.  This is unexpected because it is natural to assume that reducing the distance between the cavity and waveguide should improve the overlap integral between the two modes.  However, this increase in the quality factor is observed under a large variety of different simulation parameters, suggesting that it is real, as opposed to a simulation artifact.  We suspect that, at four hole separation, the anti-node of one of the modes overlaps with the node of the other resulting in an lower overlap integral.  Further investigation is required in order to conclusively confirm this.

The coupling changes from {$ Q _{wg} \approx 500 $} for the tilted cavity with two-hole separation to {$ Q _{wg} \approx  10^6 $} for four and five-holes separation (both configurations). For single photon sources based on PC cavities with InGaAs QDs operating at {$ 900 nm - 1000 nm$}, the experimental out-of-plane quality factor is limited to about {$ Q _{c} = 10^4 $} because of material loss and fabrication imperfections \cite{Englund05_OptExp}. On the other hand, to get efficient photon transfer into the waveguide, {$ Q _{wg} $} needs to be lower than {$ Q _{c} $} therefore, only the coupling configurations with two- and three-hole separation represent good options. Experimentally we expect the total Q to be independent of the waveguide coupling in the case of four and five holes separation.

To test the validity of our simulation results, the couplers were fabricated on a 165 nm thick freestanding GaAs membrane containing a InGaAs QD layer. Structures with two- to five-hole separation in both tilted and straight configuration (Fig. 1(c, d)) were fabricated. We made seven structures of each kind. The spectrum of each cavity was measured using the InGaAs QDs embedded in the GaAs membrane as an internal light source. The fabrication and measurement procedures are similar with those reported in \cite{05PRLEnglund}. The mean value of the quality factor for each configuration is plotted in Fig. 3(a), where the error bars are given by the standard deviation in Q due to fabrication fluctuations between the seven structures of each kind.


\begin{figure}[htbp]
    \includegraphics[width=3.5in]{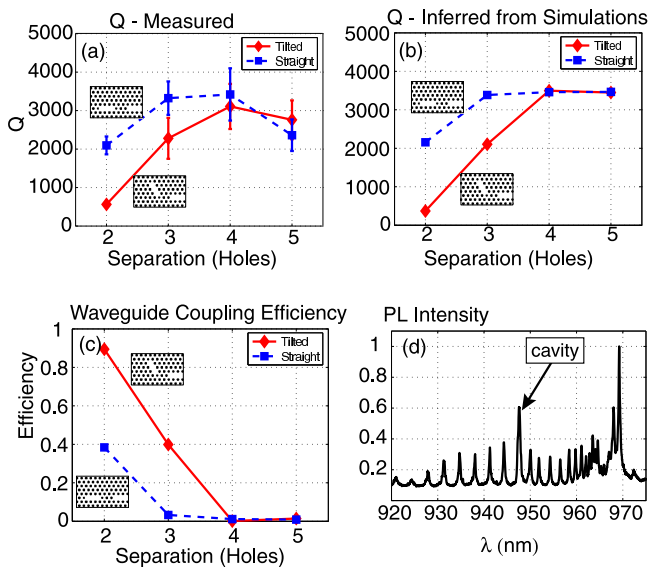}
    \caption{ Comparison between simulations and experimental data for cavity-waveguide couplers. (a) The measured value of total Q (mean) (b) The value of Q inferred from simulations by combining simulated {$Q _{wg}$} and measured {$Q _{c}$}.  (c) The coupling efficiency from the PC cavity into the PC waveguide. (d) Measured spectrum of a closed waveguide coupled to a L3 cavity. The Fabry-Perot fringes are equidistant in the linear region of the waveguide dispersion relation (where the cavity is also located) and they get closer next to the waveguide band-edge (970nm).}
\end{figure}

As expected from simulations, the experimental data show that for the same cavity-waveguide separation, the total quality factor is lower for the tilted than for the straight configuration. This result is a consequence of higher coupling for tilted cavities.

Since a more efficient coupling between the cavity and the waveguide degrades the cavity quality factor, when designing a PC network one should choose the configuration that gives the optimum trade-off between transfer efficiency and high Q. One advantage of using the tilted cavity is that the same set of parameters can be obtained with the cavity further spaced from the waveguide. 

As mentioned before, the cavity coupling was designed to couple in the linear region of the waveguide-band dispersion relation. To test the position of the cavity with respect to the waveguide band, we fabricated longer waveguides closed at the ends. These waveguides act as Fabry-Perot resonators. Fringes can be observed using the broad distribution of the QDs \cite{viktorovitch}. In the linear region of the dispersion relation the fringes are equally spaced, and get closer together as the frequency approaches the band edge. Since the cavity resonance was positioned in the region with equidistant fringes, we concluded that the coupling occurs in the linear region (Fig. 4(d)).

For a direct comparison between simulation and experiment, {$ Q _{c}$} of the uncoupled cavity needs to be known. The upper bound for {$ Q _{c}$} is limited by fabrication imperfections and material loss. Our simulation results indicate that in the case of coupled cavities with four hole separation the coupling into the waveguide is very small so the total Q is well approximated by {$ Q _{c}$}. For this reason, the average value of the measured Q for the tilted configuration with four hole separation was used as {$ Q _{c}$}. By plugging {$ Q _{c}$} and the simulated value for {$ Q _{wg}$} into expression (1), the predicted value for the total Q ({$ Q _{tot}$}) was computed. The values for the Q inferred from simulations are plotted in Fig. 3(b) and show good agreement with the experimental data (Fig. 3(a)). Some inconsistency is observed in the case of five-hole separation. These inconsistencies result from fabrication errors.

The coupling efficiency into the waveguide was computed by taking the ratio $Q / Q_{wg}$ and the results are plotted in Fig. 3(c). The coupling efficiency is up to 90\%  in the case of tilted configuration with two holes separation and up to 40\% for straight configuration with two holes separation.

In conclusion we have designed PC cavity-waveguide couplers with optimized coupling efficiency and operating in the linear waveguide dispersion region. We have shown both theoretically and experimentally that the coupling between a L3 PC cavity and PC waveguides can be improved by tilting the cavity with respect to the waveguide. The coupling is more efficient because the evanescent tails of the cavity field are not oriented along the cavity axis but at a {$ 30^{o} $} angle. Understanding and controlling the coupling mechanism is essential for on-chip single photon transfer and the implementation of on-chip quantum networks.

Financial support was provided by the MURI Center for photonic quantum information systems (ARO/DTO program No. DAAD19-03-1-0199), ONR Young Investigator Award and NSF Grant No. CCF-0507295.

\bibliographystyle{unsrt}
\bibliography{references}

\end{document}